\documentclass[epj]{svjour}
\usepackage{graphics}
\usepackage{amssymb}
\begin{document}
\title{Monte Carlo simulations of the classical two-dimensional discrete frustrated $\phi ^4$ model}
\author{V.V. Savkin\inst{1} \thanks{\emph{e-mail:} vladimir@shg.ru} \and A.N. Rubtsov\inst{1} \and T. Janssen\inst{2}
%
}                     
%
%
\institute{Physics Department, Moscow State University, 119992
Moscow, Russia \and Institute of Theoretical Physics, University
of Nijmegen, Postbus 9010, 6500 GL Nijmegen, The Netherlands}
\date{Received: date / Revised version: date}
%
\abstract{ The classical two-dimensional discrete frustrated $\phi
^4$ model is studied by Monte Carlo simulations. The correlation
function is obtained for two values of a parameter $d$ that
determines the frustration in the model. The ground state is a
ferro-phase for $d=-0.35$ and a commensurate phase with period
$N=6$ for $d=-0.45$. Mean field predicts that at higher
temperature the system enters a para-phase via an incommensurate
state, in both cases. Monte Carlo data for $d=-0.45$ show two
phase transitions with a floating-incommensurate phase between
them. The phase transition at higher temperature is of the
Kosterlitz-Thouless type. Analysis of the data for $d=-0.35$ shows
only a single phase transition between the floating-fluid phase
and the ferro-phase within the numerical error.
} 
\maketitle

\section{Introduction}
\label{intro}

One of the microscopic origins of the spatially modulated
structures is given by the competing interactions between the
particles. The Frenkel-Kontorova model, the axial
next-nearest-neighbour Ising (ANNNI) model, and the discrete
frustrated $\phi ^4$ (DIFFOUR) model provide basic microscopic
Hamiltonians for the systems of this class. These models have been
studied extensively, specially the first two~\cite{ANNNI1,ANNNI2}.
Competing interactions lead to the appearance of incommensurate
and commensurate regions in the phase diagrams of these models.
The order parameter in these systems is a vector that determines
the amplitude and the phase of modulation. For the incommensurate
state, the free energy is degenerate in the phase of modulation.
The two-dimensional case is of special interest here, since there
is no long-range order in such planar systems if the order
parameter has more than one component~\cite{PP}. A nonmonotonic
algebraic decay of correlations appears instead of the usual
incommensurate state~\cite{ANNNI2}.

The correlation function in two-dimensional systems with competing
interactions has at least three possible types of behaviour, as
found in the spin models~\cite{ANNNI2}. The disordered phase shows
an exponential monotonic or modulated ($q\neq 0$) fall-off of the
correlation function:
\begin{equation}
C(r)=<S(0)S(r)> \sim \exp(-r/\xi) \cos(q r). \label{eq000}
\end{equation}
The case $q\neq 0$ is usually called floating-fluid phase. An
algebraical decay of correlations at $r \rightarrow \infty$ is a
property of the floating-incommensurate (FIC) or floating-solid
phase:
\begin{equation}
C(r) \sim r^{-\eta} \cos(q r + \varphi). \label{eq00}
\end{equation}
Finally, a commensurate phase with locked-in wavevector $q_0$ is
not destroyed by the fluctuations in two-dimensional case.
Therefore the correlation function in this case takes the form
\begin{equation}
C(r) \sim \cos(q_0 r). \label{eq0}
\end{equation}
The commensurate case includes the ferro-phase also ($q_0=0$).

One should note that in the two-dimensional case the phase
transition from the para-phase to the FIC-phase is believed to be
related with the formation and dissociation of the dislocations or
vortices~\cite{ANNNI2}. This scenario of a phase transition can be
described by the Kosterlitz-Thouless theory which is basic for the
two-dimensional $XY$ universality
class~\cite{KT,Pelissetto,Gupta}. The power law dependence of
$C(r)$ for this universality class appears everywhere below the
critical temperature, down to $T=0$ or to another phase transition
(to a commensurate state, for example). The critical index $\eta$
is equal to $0.25$ in the point of the phase transition ($T_c$)
and decreases for decreasing temperature as
$\eta(T)=0.25-C\sqrt{T_c - T}$, where $C$ is a constant ($\eta
\sim T$ for $T \rightarrow 0$). The thermal dependence of the
correlation length above the transition point is of the form $\xi
\sim \exp(b/\sqrt{T-T_c})$, where $b$ is a constant.

One of the well studied spin models with competing interactions is
the ANNNI model. The ANNNI model is a spin lattice model with $\pm
1$ classical spins, ferromagnetic nearest-neighbours interaction
$J_1$ in all directions and antiferromagnetic
next-nearest-neighbour interaction $J_2$ in a single direction.
The ANNNI model in two dimensions has been studied by numerous
methods~\cite{ANNNI1,ANNNI2}. The ground state of this model at
zero temperature is a ferromagnetic phase for $-J_2/J_1<0.5$ and a
$<2>$ phase (two spins up and two spins down $\upuparrows
\downdownarrows$) for $-J_2/J_1>0.5$. Obviously, an FIC-phase can
appear only at finite temperature between the paramagnetic phase
and the $<2>$ phase for $-J_2/J_1>0.5$. The behaviour of the model
was studied by transfer-matrix calculations and
finite-size-scaling analysis~\cite{Beale}, Monte Carlo
simulations~\cite{Selke,Sato}, dynamical Monte Carlo
method~\cite{Barber}, fermion approximation~\cite{Villain} and its
modification~\cite{Grynberg}, the interface free energy
method~\cite{Muller}. Some of these results seem to indicate a
Kosterlitz-Thouless~\cite{KT} type of the phase transition from
the paramagnetic to the FIC-phase and a
Pokrovsky-Talapov~\cite{PT} type of phase transition from this
phase to $<2>$. The latter scenario implies a square-root power
law change of the wavevector with respect to temperature near the
transition to the $<2>$ phase. One should point out that the
one-dimensional quantum ANNNI model studied in~\cite{Rieger} also
reveals all phases and types of the phase transitions mentioned
above. The approximations give a spread of the phase transition
temperature over a rather wide region. On the other hand, other
calculations predict an FIC-phase with a width comparable to that
spread. The recent results on a cluster heat bath
simulation~\cite{Sato} of systems of about $64 \times 128$ atoms
demonstrate that the FIC region is at least 2 times smaller than
the previous papers predicted~\cite{Beale}. Further, the
nonequilibrium relaxation method~\cite{Shirahata} for $6400 \times
6400$ systems does not predict any area for the FIC-phase at all,
so that even the presence of this phase was questioned.

In this paper we consider the classical two-dimensional DIFFOUR,
which has continuous variables and competing interactions in one
direction. The DIFFOUR model was introduced in~\cite{TJ} as a
displacement translationally invariant model for the
incommensurate phases. Instead of $\pm 1$ spins, this model
considers double-well anharmonic 2-4 oscillators at the nodes of a
lattice. Similarly, there is a harmonic nearest-neighbour and
next-nearest-neighbour coupling. The potential energy can be
written as
\begin{eqnarray}
V &=& - \frac{a}{2}\sum\limits_i {u_i^2}+ \frac{a}{4}
\sum\limits_i {u_i^4} +\frac{1}{2} \sum\limits_{i,j} (u_i - u_j)^2
\sigma_{ij} \nonumber \\ &+&\frac{d}{2} \sum\limits_{i,j} (u_i -
u_j)^2 \sigma'_{ij}. \label{eq1}
\end{eqnarray}

Here $\sigma_{ij}=1$ for the nearest neighbours and
$\sigma_{ij}=0$ otherwise, while $\sigma'_{ij}=1$ for the
next-nearest-neighbours in the direction $z$ (see Fig. 1a). The
phase diagram for the analogous three-dimensional model is
presented in~\cite{AR}. The system has commensurate and
incommensurate regions in the phase diagram for $d<0$ i.e. with
competing interactions.

The ANNNI model can be considered as a limiting case of the
DIFFOUR model ($a\rightarrow +\infty$). It corresponds to the
order-disorder limit of the latter model. In the other limit
($a\rightarrow +0$) of the DIFFOUR model the ground state may be
an incommensurate phase which is reached via a displacive
transformation~\cite{TJ1}. Furthermore, the greater freedom in the
DIFFOUR model could also possibly lead to the existence of a
broader the FIC-phase than in the ANNNI model. In view of the
unclear situation in the latter we want to study the existence of
FIC-phase in the DIFFOUR model. We shall do this using Monte Carlo
simulations. The mean-field phase diagram can easily be determined
and is presented in Fig. 1b.

\begin{figure*}
\includegraphics{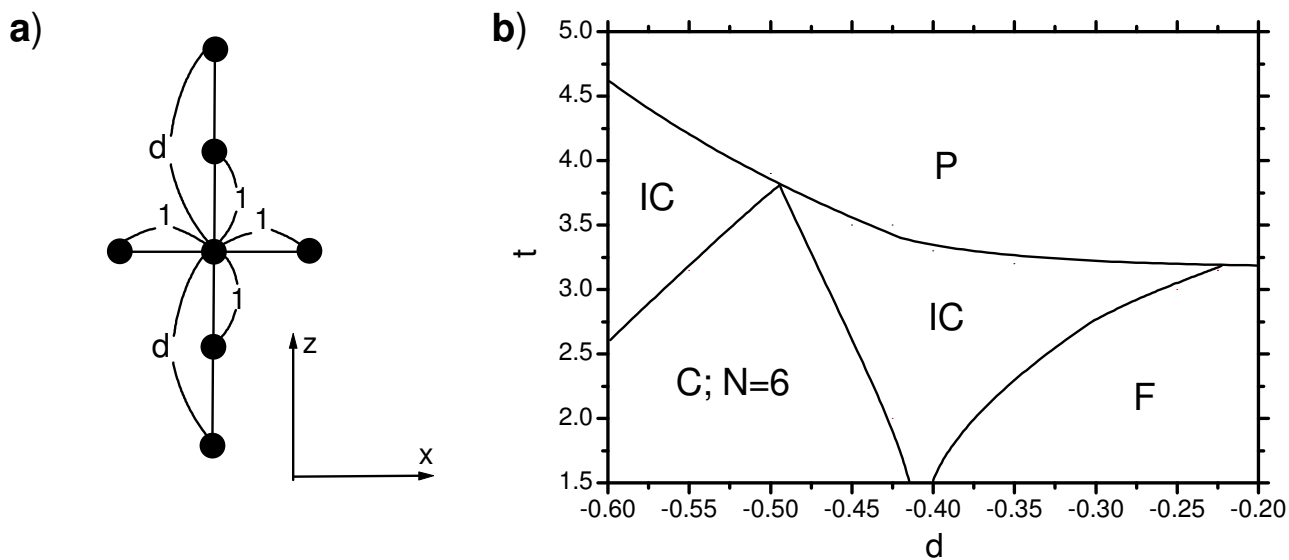}
\caption{a) Scheme of interactions in the two-dimensional DIFFOUR
model. b) Mean-field phase diagram of the two-dimensional DIFFOUR
model for the parameter $a=5$ ($t$ is the temperature hereafter).
P - para-phase, F - ferro-phase, IC - incommensurate phase, C -
commensurate phase. There are also narrow regions of the
commensurate phases in the area of the incommensurate phase not
shown in the picture.}
\label{fig:1}       
\end{figure*}

Correlation functions of the model are analyzed for certain
parameters; their behaviour is always one of the three scenarios
mentioned above (\ref{eq000},\ref{eq00},\ref{eq0}). An FIC-phase
has been found. Indications of the Kosterlitz-Thouless scenario of
the phase transition from the floating-fluid to the FIC-phase are
discussed. The transition to the ferro-phase occurs via the
floating-fluid phase. Surprisingly, the period of this phase {\it
increases} as temperature decreases.

\section{Algorithm}
\label{sec:2}

We study the two-dimensional DIFFOUR model using the Monte Carlo
method. The difficulty of the simulation of the models with
competing interactions is in the complexity of their configuration
space which usually has many deep almost degenerate local minima.
Then, it is difficult to overcome a potential barrier and to
change the period of the modulation. This situation can lead to an
unacceptably long computation time using conventional Monte Carlo
schemes, particularly for large systems. Special algorithms should
be used to achieve a reasonable result ~\cite{Sato,Shirahata}.

In this paper, we use the "local-heating" algorithm which worked
well for the three-dimensional DIFFOUR model ~\cite{AR}. The idea
is to modify the potential energy in certain regions of the system
to allow an easier switch between local minima. Actually, the
temperature of the system is not changed but the constants of the
potential energy are modified. However, for the largest part of
the system its potential energy is still determined by
(\ref{eq1}). The regions with a modified potential energy are not
included in the calculation of the physical quantities (order
parameter, correlation functions), while the usual Metropolis
algorithm for Monte Carlo sampling is applied to the whole system.

We use periodic boundary conditions and the following modification
of the potential energy:
\begin{eqnarray}
V &=& - \frac{a}{2}\sum\limits_i \theta_i {u_i^2}+ \frac{a}{4}
\sum\limits_i \theta_i {u_i^4} + \frac{1}{4}\sum\limits_{i,j}
(\theta_i+\theta_j) (u_i-u_j)^2 \sigma_{ij} \nonumber
\\ &+& \frac{d}{4}
\sum\limits_{i,j} \theta_{(i+j)/2} (u_i-u_j)^2 \sigma'_{ij},
\nonumber \\    \theta_i &=& (1+l_1 (
\exp[-(x_i^2+z_i^2)/l^2_2]\nonumber\\
&+&\exp[-((N_x-x_i)^2+(N_z-z_i)^2)/l^2_2]))^{-1}, l_1>0.
\label{eq2}
\end{eqnarray}
$N_x$ and $N_z$ are the sizes of the lattice along the $x$ and
$z$-directions, respectively. The new coefficients $\theta_i$
(\ref{eq2}) mean that their values in the vicinity of the lines
$x_i=0$ and $z_i=0$ are smoothly decreased. Parameters $l_1,l_2$
determine the height and the width of this decrease, respectively.
The change of the parameters mimics a local heating of the
corresponding regions. The idea of this procedure is to put these
parts of the system in a para-phase and to destroy the local
modulation phase. If the profile $\theta(x,z)$ is smooth enough,
the phases of modulation at the two sides of the heated line are
in an ideal case independent and can randomly fluctuate. The whole
system may change the period of the modulation now. Use of "local
heating" leads to faster thermalization of the system.

Additionally we realised that it is useful to simulate the systems
with a larger size along the $x$ direction. The possible
explanation is that once the period is changed at some $x$, it is
relatively easy for this change to spread to the whole system. The
probability for this initial change is proportional to the
$x$-size.

We calculate the correlation function as follows:
\begin{equation}
C(\vec{r})=<u(\vec{r}')u(\vec{r}'+\vec{r})>|_{\vec{r}'}.
\label{eq3}
\end{equation}
Here $<>|_{\vec{r}'}$ means an ensemble average. We consider two
functions $C_x=C(\vec{r}=r_x)$ (along $x$ axis) and
$C_z=C(\vec{r}=r_z)$ (along $z$ axis) hereafter. The presence of
the heated regions is taken into account as follows. We add data
to the point $\vec{r}$ of correlation function (\ref{eq3}) if
there is no "local heated" region between $\vec{r}$ and $\vec{r}'$
and add data to the point $\vec{r}-\vec{r}'$ otherwise. If one of
the points $\vec{r}$ or $\vec{r}'$ appears to be in the "local
heated" region the data are not included in the calculation of the
correlation function. This procedure reduces the influence of the
finite-size effects. The data for the correlation function appear
to be reliable even for distances larger than half the system
size.

We check whether the approach of the "local heating" algorithm by
changing the size of the system and by starting with various
initial distributions of atom's displacements is reasonable. One
should use large enough sizes of the system since it is difficult
to distinguish exponential and power law decay of correlations.
The situation becomes worse when we deal with modulated structures
with large modulation period. We found that an optimal size for
the present algorithm is about $200\times 100$. The total number
of Monte Carlo samplings for the given parameters of the model
$a,d$ and temperature $t$ is about $10^{10} \div 10^{11}$.
Parameters of "local heating" are varied in a region $2 \lesssim
l_1,l_2 \lesssim 4$. We have checked that a two-times increase of
the slab size in the $x$ or $z$-direction does almost not change
the exponents of power or exponential decaying or the period of
the modulation within the numerical accuracy. The result does not
depend on the initial configuration of the system. Thus, we
believe that the presented algorithm allows us to perform accurate
numerical simulations.

\section{Results and their analysis}
\label{sec:3}

We present here detailed results for two values of the parameter
$d$ at $a=5$: $d=-0.45$ and $d=-0.35$. We choose these values of
the parameter $d$ since at least the mean-field approximation
predicts various ground states for zero temperature (commensurate
phase with $N=6$ for $d=-0.45$ and ferro-phase for $d=-0.35$, see
Fig. 1b). The parameter value $a=5$ of the DIFFOUR model
(\ref{eq1}) is used since the amplitude of modulation then has a
value comparable with the value of $C(0)$. It makes the analysis
of the data easier. We have found that the modulation amplitude is
much less than the value $C(0)$ for larger values of $a$ (for
example, at $a \gtrsim 20$). The mean-field approach for the
DIFFOUR model confirms these observations.

The correlation functions $C_z$ and $C_x$ are calculated at
several values of the temperature. The results for $C_z$ are
presented in Fig. 2 and Fig. 3 for $d=-0.45$ and $d=-0.35$,
respectively. Typical dependencies of $C_x$ for both values of the
parameter $d$ are shown in Fig. 4. We use a log-log scale for the
$C_x$ functions. The most informative dependencies here are $C_z$
(Fig. 2,3). Qualitatively the behaviour of the functions $C_x$ and
$C_z$ is different: non-monotonic and monotonic dependence on the
distance, respectively. We believe that the influence of the
finite-size effects starts at $x \gtrsim 100$ and at $z \gtrsim
70$ and we do not consider these regions in the analysis of the
data. The error bar for $C_x$ and $C_z$ at moderate values of
temperature for $d=-0.45$ and $d=-0.35$ is about a few percent and
becomes larger at lower temperatures.

We analyze the obtained results by fitting the data for $C_z$
(Fig. 2,3) using one of the functions (\ref{eq000},\ref{eq00}). We
also try to find a power law decay for $C_x$ (Fig. 4) if it is
possible.

\begin{figure*}
\includegraphics{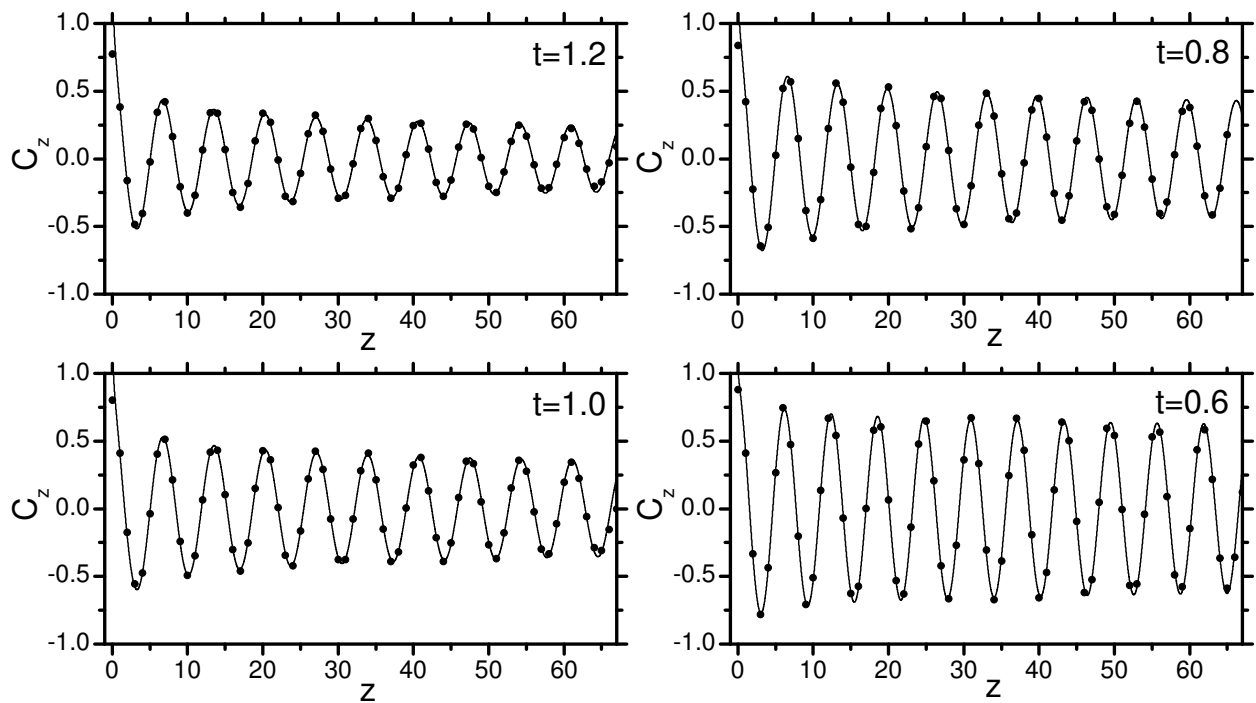}
\caption{The correlation function $C_z$ for various temperatures:
$t=1.2$, $t=1.0$, $t=0.8$ and $t=0.6$. Parameters of the DIFFOUR
model are $a=5$, $d=-0.45$. Dots - Monte Carlo results, solid
lines - results of fitting by non-monotonic power law decay
(\ref{eq00}). The Kosterlitz-Thouless like phase transition takes
place at $t_1 \approx 1.2$ (index $\eta=0.25$).}
\label{fig:2}       
\end{figure*}

\begin{figure*}
\includegraphics{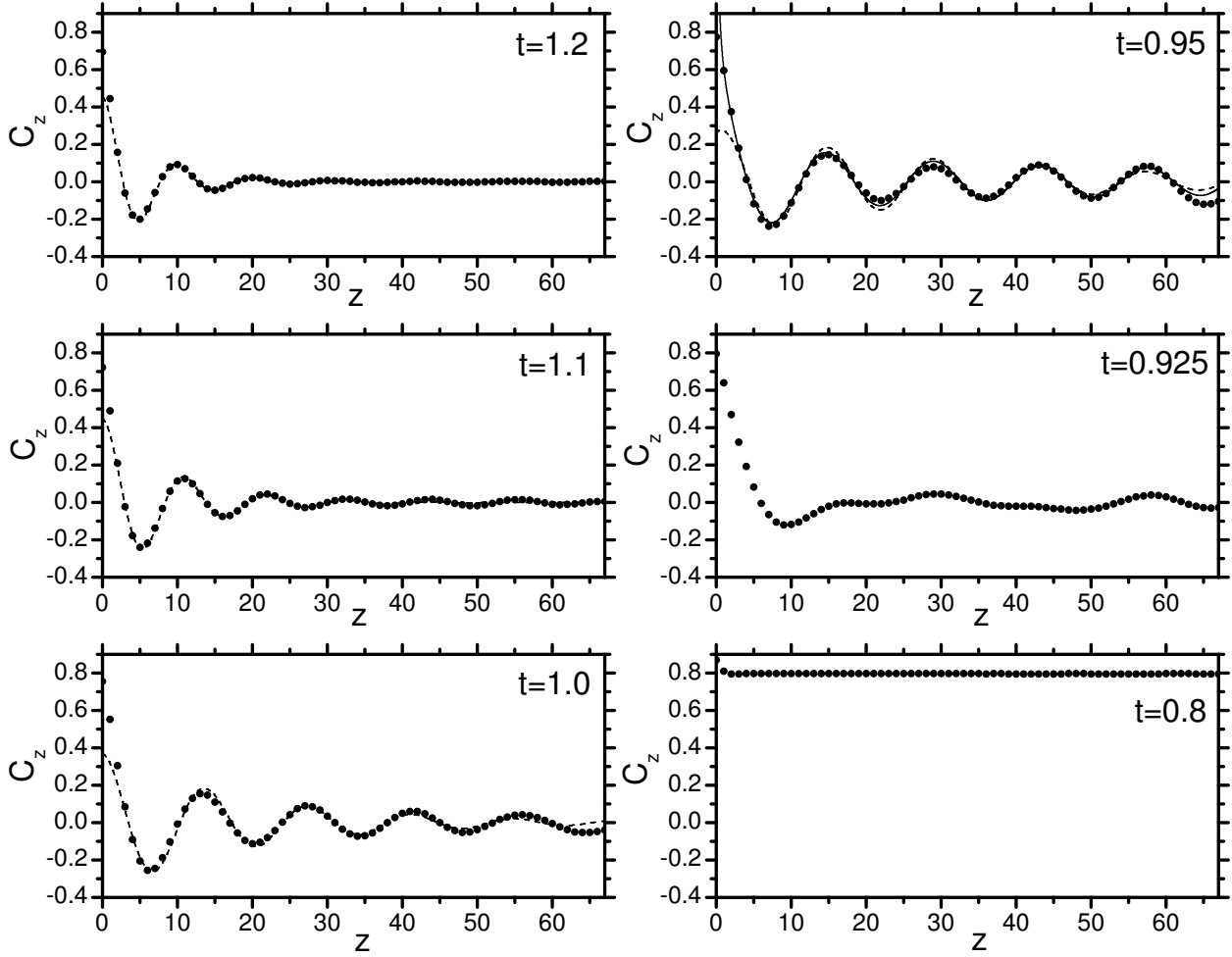}
\caption{The correlation function $C_z$ for various temperatures:
$t=1.2$, $t=1.1$, $t=1.0$, $t=0.95$, $t=0.925$ and $t=0.8$.
Parameters of the DIFFOUR model are $a=5$, $d=-0.35$. Dots - Monte
Carlo results, dashed lines - results of fitting by non-monotonic
exponential decay (\ref{eq000}). Solid line - result of the
fitting at $t=0.95$ by non-monotonic power law decay (\ref{eq00}):
index $\eta$ is approximately equal to $0.5$. Single phase
transition from floating-fluid phase to the ferro-phase takes
place at $t_c \approx 0.91$.}
\label{fig:3}       
\end{figure*}

\begin{figure*}
\resizebox{2.0\columnwidth}{!}{
\includegraphics{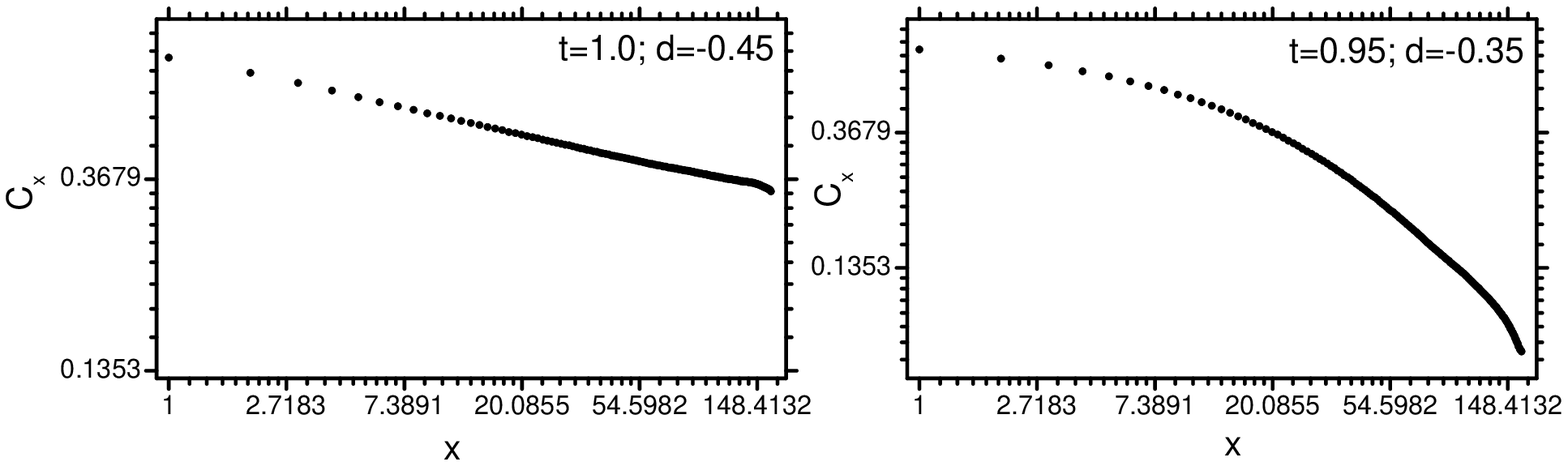}}
\caption{Typical dependencies of the correlation function $C_x$
for $d=-0.45$ and $d=-0.35$ (temperatures are equal to $t=1.0$ and
$t=0.95$, respectively). Logarithmic scale is used for both axes.
Results are obtained by Monte Carlo simulations.}
\label{fig:4}       
\end{figure*}

\subsection{Case $d=-0.45$}
\label{sec:31}

We start with the case $d=-0.45$ (Fig. 2). We look for the
possibility of a phase transition from the para-phase to the
FIC-phase. The dependencies of the correlation functions $C_z$ at
$t \gtrsim 1.2$ can be fitted by equation (\ref{eq000}), i.e. by
non-monotonic exponential decay. There is no power law decay for
$C_x$ either for these values of temperature. Thus, we conclude
that it is floating-fluid phase at $t>1.2$ and $d=-0.45$.

One can fit the dependencies $C_z$ in the vicinity of the $t \sim
1.2$ by nonmonotonic power law decay (\ref{eq00}). The results of
this fitting are presented in Fig.2 by solid lines. It is possible
also to find a power law decay for $C_x$. This region is believed
to be close to the possible point of the phase transition from the
floating-fluid phase to the FIC-phase. The value of the index
$\eta$ is approximately equal to $0.25$ at the temperature $t_1
\approx 1.2$. This coincides with the condition of the
Kosterlitz-Thouless phase transition. One can find, that data for
$C_z$ (Fig. 2) for the region $0.55<t<1.2$ can also be fitted by a
non-monotonic power law dependence (\ref{eq00}) with index $\eta$
depending on temperature. We plot the dependence of the
$(0.25-\eta(t))^2$ on $t_1 - t$ for the $t<1.2$ (Fig.5a). One sees
a linear dependence, which is a feature of the Kosterlitz-Thouless
type behaviour~\cite{KT,Gupta}. Thus, the phase transition from
the floating-fluid phase to the FIC-phase at $t_1 \approx 1.2$ is
believed to be of Kosterlitz-Thouless type. The data for $C_x$ at
$0.55<t<1.2$ can also be fitted by a monotonic power law decay.
Typical dependence of $C_x$ for this temperature region is
presented in Fig.4 ($t=1.0$). In our calculations, the indices
$\eta$ for $C_x$ and $C_z$ do not coincide (the difference is
about $10-20\%$). Although, we don't know whether this difference
is of physical or numerical origin, we rather attribute it to
finite-size effects in simulations. Nevertheless, we use here the
index $\eta$ obtained from the data of fitting in the $z$
direction, i.e. by equation (\ref{eq00}) for analysis of the data.

We also show in Fig. 5b how the wavevector changes as a function
of temperature. We find that the modulation period is equal to
$N=6$ at $t\leq 0.55$. The appearance of the domain walls
(discommensurations - small regions with pseudo-period $N \neq 6$)
can be seen at $t\gtrsim 0.55$. It prevents the system to get a
commensurate modulation period ($N=6$). The point $t_2 \approx
0.55$ is supposed to be the point of the phase transition from the
FIC-phase to the commensurate phase with period $N=6$ (see Fig.
5b). This picture is similar to that for the two-dimensional ANNNI
model~\cite{Sato} for the transition from the FIC-phase to the
$<2>$ phase. One should note that the accuracy of our results is
not sufficient to determine whether the phase transition from
FIC-phase to the $<2>$ phase is of Pokrovsky-Talapov type or not
(Fig. 5). Present results are in agreement with previous
conclusions by Schulz and Haldane {\it et al} that only
commensurate phase with sufficiently small periodicity can
transform into a FIC-phase~\cite{Schulz,Haldane}. It is also
possible that narrow commensurate regions exist in the temperature
range $0.55<t<1.2$. Study of this problem demands more accurate
methods of numerical simulations and analysis of the data.
\begin{figure*}
\resizebox{2.0\columnwidth}{!}{
\includegraphics{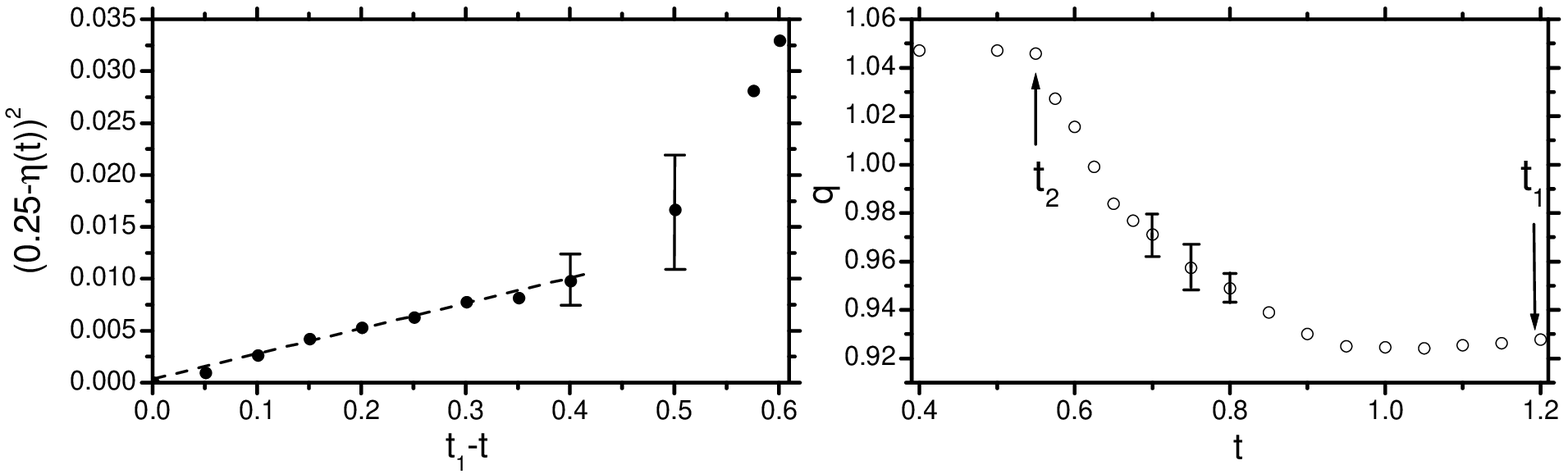}}
\caption{a) Dependence of the function $(0.25-\eta(t))^2$ on the
$t_1-t$ below the point of the phase transition from
floating-fluid phase to the FIC-phase (parameters of the model
$a=5$, $d=-0.45$). Dots are the results of the analysis of the
Monte Carlo data (Fig. 2), dashed line is a linear approximation
in the vicinity of the phase transition. b) Dependence of the
wavevector $q$ on the temperature $t$ (parameters of the model
$a=5$, $d=-0.45$). Dots are the results of the analysis of the
Monte Carlo data (Fig. 2). $t_1$ and $t_2$ are the temperatures of
the phase transitions from the floating-fluid phase to the
FIC-phase and from the FIC-phase to the commensurate phase with
period $N=6$, respectively.}
\label{fig:5}       
\end{figure*}

\subsection{Case $d=-0.35$}
\label{sec:32}

Before presenting the numerical results, let us stress several
features of the $d=-0.35$ situation. The mean-field predicts two
transitions: from the para-phase to an incommensurate one, and
than to the ferro-phase. The three-dimensional DIFFOUR model
indeed behaves so at certain parameters~\cite{AR}. On the other
hand, the ANNNI model does not show two transitions; there is only
a single transition from the floating-fluid (disordered) to the
ferro-phase. For the present system, because of the relatively
small value $a=5$, the modulation is much more pronounced and
falls off slower than in the ANNNI model. As well, the mean-field
region for the incommensurate phase is quite wide. Therefore one
could expect a transition from a floating-fluid to an FIC-phase
for $d=-0.35$. Surprisingly, we did not found evidence for this;
qualitatively the result coincides with that of the ANNNI model.

Results for the case $d=-0.35$ for $C_z$ are presented in Fig. 3.
Data for $t\gtrsim 1.0$ can be fitted by non-monotonic exponential
decay (\ref{eq000}). Results of fitting $C_z$ are shown in Fig. 3
by dashed lines. We have performed a detailed studies of the
region $1.0 < t < 0.8$ (part of them is shown in Fig. 3). Since
the period of the modulation is more than $14$ in this region it
is difficult to distinguish whether there is a non-monotonic power
law decay or not. Besides, the fluctuations in this region are
increased and our algorithm does not allow us to judge the type of
decay of the correlations. We try to fit these dependencies by
exponential and power law decay. The results for $t=0.95$ are
presented in Fig. 3 by dashed and solid lines, respectively. Both
dependencies show agreement with Monte Carlo results within the
numerical error. The index $\eta$ for power law decay is found to
be about $0.5$. Thus, we believe that there is no
Kosterlitz-Thouless phase transition for $d=-0.35$ since there is
no agreement with a power law decay (\ref{eq00}) with the exponent
value $\eta=0.25$ in this temperature region. One can see that the
period of the modulation is increased, while the amplitude of
modulation is decreased. This situation is opposite to the case
$d=-0.45$. A typical result for the $C_x$ is presented in Fig. 4
($t=0.95$). It is difficult to find power law decay in the region
$1.0 < t < 0.8$ for $C_x$. Certainly, there is no index
$\eta=0.25$ in this case either. It is difficult to obtain clear
dependence of correlations decay in the vicinity of the phase
transition point due to increase of fluctuations and strong
influence of the finite-size effects (see Fig. 3, $t=0.925$).
Thus, we conjecture that there is only one second order phase
transition for $d=-0.35$ from the floating-fluid phase into the
ferro-phase. We estimate the temperature of this phase transition
to be $t_c \approx 0.91$. We would like to note that one can
suggest a transition from floating-fluid phase to a ferro-phase
via a phase with monotonic exponential decay of correlations with
$q=0$ (\ref{eq000}) and the existence of a disorder line by
analogy with the ANNNI model~\cite{Beale}. This phase may exist in
a very narrow temperature region at given parameters of the model
and data for $t=0.925$ confirm this. The question of the presence
of a disorder line in the phase diagram of the two-dimensional
DIFFOUR model needs an additional study.

\section{Conclusions}
\label{sec:4}

In conclusion, we have studied the DIFFOUR model in two dimensions
by Monte Carlo simulations. The classical DIFFOUR model has two
parameters $a$ and $d$, where the latter one is the effective
constant of the competition of the interactions. Two sets of
parameters of the model have been considered here in detail: $a=5,
d=-0.45$ and $a=5, d=-0.35$. A "local heating" algorithm has been
used for numerical simulations. This algorithm uses modifications
of the constants and parameters of the potential energy of the
DIFFOUR model. Correlation functions in both directions have been
calculated for analysis of the behaviour of the model.

We have found that correlation functions for the $a=5, d=-0.45$
can be fitted by a non-monotonic exponential decay at $t>1.2$ and
by non-monotonic power law decay at temperatures $0.55<t<1.2$. The
phase transition at $t_1 \approx 1.2$ of Kosterlitz-Thouless type
is believed to take place with $\eta=0.25$ from the floating-fluid
phase to the FIC-phase. The dependence of the index $\eta(t)$
below the point of the phase transition has a square root of
$t_1-t$ behaviour in agreement with the Kosterlitz-Thouless
theory. The ground state of the model at $t<0.55$ ($a=5, d=-0.45$)
is a commensurate phase with modulation period $N=6$. The second
phase transition at these parameters of the model from FIC-phase
to the commensurate one occurs at $t_2 \approx 0.55$.

The correlation functions for the case $a=5, d=-0.35$ can be
fitted by a non-monotonic exponential decay for $t>0.95$.
Dependencies at $t \sim 0.95$ can be fitted by non-monotonic power
law decay also but with $\eta>0.25$. Numerical data do not allow
us to distinguish exponential from power law decay for this case
since the period of modulation is too large. Nevertheless, we
believe that there is no FIC-phase in this case and only one phase
transition from the floating-fluid phase to the ferro-phase, at
$t_c \approx 0.91$. The ground state of the model for $t<0.9$ is a
ferro-phase.

We would like to point out that the results of the Monte Carlo
simulations (Fig. 2,3) and the mean-field data (Fig. 1) are
different. Mean-field approximation predicts an incommensurate
phase, while Monte Carlo results show a FIC-phase with power law
decay of correlations. Mean-field theory predicts two phase
transitions for $d=-0.35$ and a region of the incommensurate
phase, while Monte Carlo simulations show only one phase
transition from the floating-fluid phase to the ferro-phase.
Nevertheless, the low-temperature phase (ferro-phase for $d=-0.35$
and commensurate phase with period $N=6$ for $d=-0.45$) is
predicted correctly in the mean-field approximation, as in the
limit $t=0$ fluctuations are absent and the mean-field result is
exact.

\begin{acknowledgement}
This work was supported by INTAS project for young scientists
(grant YSF 2001/1 - 135).
\end{acknowledgement}

\end{document}